\documentclass[12pt]{iopart}

\usepackage{epsfig}

\expandafter\let\csname equation*\endcsname\relax
\expandafter\let\csname endequation*\endcsname\relax

\usepackage{amsmath}

\newcommand{\ket}[1]{|#1\rangle}

\begin{document}

\title{Quantum noise eater for a single photonic qubit}

\author{Miroslav Gavenda, Lucie \v{C}elechovsk\'{a}, Miloslav Du\v{s}ek and Radim Filip}

\address{Department of Optics, Faculty of Science, Palack\' y University, 17.
listopadu 1192/12,  77146 Olomouc, Czech Republic}

\date{\today}

\begin{abstract}
We propose a quantum noise eater for a single qubit and experimentally verify
its performance for recovery of a superposition carried by a dual-rail
photonic qubit. We consider a case when only one of the rails (e.g., one of interferometric arms) is vulnerable to noise. A coherent but randomly arriving photon penetrating into
this single rail causes a change of its state,
which results in an error in a subsequent quantum information
processing. We theoretically prove and experimentally demonstrate a
conditional full recovery of the superposition by this quantum noise eater.
\end{abstract}


\maketitle

\section{Introduction}

A success of any application of quantum physics strongly depends on
accessibility and quality of quantum resources. Quantum bit (qubit), being
quantum analog of classical bit, is a fundamental but fragile element of
quantum information \cite{niels01}. It is the simplest quantum system with
the smallest Hilbert space of quantum states consisting all possible
superpositions of two basis states. Many physical systems have been
experimentally proved to exhibit such superpositions applicable for
quantum information processing
\cite{bouw96,marc03,lomb02,mort08,elze04,trau07,imam99,clar08,naka99,chio03,ansm09,lem11,arn12}.
However, a quantum superposition can
simply be lost by noise driving qubit to a mixture of basis states
\cite{niels01,zurek03}. The noise can exhibit many different
characteristics depending on a coupling of qubit to a noisy environment and
also on a state of the noisy environment. Many methods of protection of
single qubit against the noise have been proposed
\cite{steane96,bacon99,viola99}. Typically,
they are designed for a specific type of noise influencing well defined
Hilbert space of qubit.

However, single-qubit superposition can be also destroyed by very
destructive random coherent noise, that transforms a qubit to a system with a
higher dimension. The simplest example is a qubit represented by a (bosonic)
particle which can be coherently mixed with another
indistinguishable particle \cite{fabio09,fabio10}. It is an elementary case
of a more complex coherent continuous-variable noise \cite{weed12}, where
the number of such particles coming from the environment fluctuates. In
past, many techniques based on quantum feedback \cite{buch98} have been
experimentally verified to reduce this destructive continuous noise. A
technique commonly used in such a case is a noise eater \cite{bach03},
which is able to detect intensity of a small part of the coherent signal
mixed with noise and use adjustable feed-back loop to control the laser (or
modulate light) to reduce that noise at a cost of lower output optical
power. Recently, different techniques based on multiple copies of noisy
coherent states \cite{ander05} and measurement of noise from an environment
\cite{sab10} have been also tested. Also probabilistic version of the noise
eater reducing non-Gaussian intensity noise imposed on the coherent states
have been verified \cite{witt08}. However, a single particle ``penetrating to
a qubit'' can be even more destructive. If the signal and noise particles are
principally distinguishable but \emph{technically indistinguishable},
quality of a qubit is substantially damaged
\cite{gav11}. But even if the particles are \emph{principally
indistinguishable} and the superposition just expands coherently to a
higher dimensional state, it causes problems because many
operations are designed specifically for qubits, they expect only 2D Hilbert
space.

In this paper, we propose and perform a proof-of-principle experimental test
of the simplest coherent noise eater technique for a qubit carrying quantum
superposition. We study the simplest case when, during an elementary noise impact,
the dual-rail photonic qubit is
influenced by a single indistinguishable noise photon in only one (known) rail.
Similarly to the noise eater technique for laser light, a partial
detection of number of photons in optical beam is exploited, however with
single photon resolution. Moreover, differently to that technique, measured
information is used to herald only the cases when at most
one photon remains in the setup. To test quality of the resulting qubit, we evaluate
visibility of interference in a subsequent Hadamard gate acting on that
qubit. It appears that the proposed noise eater is able to recover visibility of
interference up to unity.

\section{Coherent noise eater for a dual-rail qubit}

The noise eater technique can be based, e.g., on
photon-number measurement which conclusively detects exactly one photon
in the propagating beam and simultaneously leaves desired superposition
unchanged. If one finds more than one photon in the beam he/she rejects that
case and does not
use the state for
further applications. 
Situations heralded by this procedure correspond to true qubits carried by individual photons. Therefore no error from multi-photon
contribution can appear.
If the noise photon is fully indistinguishable from the signal photon, the noise
effect is caused purely by an extension of the total state of the system to a higher dimensional single-mode
Hilbert space.  Such noise can be completely eliminated by the ideal noise eater.
It is an example of {\em quantum coherent}
nondestructive filtering, which requires photon number resolution.
When the signal and noise photons are \emph{partially} distinguishable (due to the different
states of non-informational degrees of freedom) the situation can be conditionally converted to the previously described ideal case by \emph{classical} filtering in front of the noise eater. Unfortunately, the required ideal nondestructive photon-number resolving
detectors are currently not feasible. 
However, we have devised an alternative implementation which
still works well for our particular situation but makes do with standard photonic technology.

Recently, a quantum relay has been used to detect whether a photon is present in a beam or not \cite{coll05,tan12}. It allows to conditionally
avoid an impact of a noise photon in a subsequent
quantum channel. However, our situation is rather opposite, since the noisy
photon could already affect the qubit. Therefore we need to end up with 
one photon exactly. In continuous variables,
coherent state filtration has been experimentally tested to avoid
non-Gaussian noise \cite{witt08}, however, it heralds on higher photon
numbers rather than on exactly a single photon state. On the other hand, in our
proof-of-principle experiment we know that only a single noise photon
can enter the system. So we can substitute nondestructive
photon-number measurement followed by rejection of any multi-photon
contributions by the {\em subtraction of a single photon} after the noise impact.
The photon is subtracted using a linear optical device that
spatially separates two incoming photons with nonzero probability and
detects one of them afterwards.
If a photon is subtracted then only one photon
remains in the dual-rail qubit. When the signal and noise photons are fully
indistinguishable, this simplified linear-optical version of the noise eater
technique can reach the perfect recovery of qubit superposition,
irrespective of the probability of the noise impact.

A single photon being in a superposition of two spatial rails can be used to
experimentally demonstrate this prospective method. Two basis states
$|0,1\rangle_{AB}$ and $|1,0\rangle_{AB}$ represent a photon being either in
the rail A or B. {\em Equatorial} states in the
form of $(|0,1\rangle_{AB}+\exp(i\varphi)|1,0\rangle_{AB})/\sqrt{2}$ carry
quantum information encoded to balanced superpositions of the basis states.
A unitary Hadamard gate is represented by symmetrical coupling between
the both rails. Ideally, it builds superposition
$|+\rangle_{AB}=(|0,1\rangle_{AB}+|1,0\rangle_{AB})/\sqrt{2}$ from
state $|1,0\rangle_{AB}$. The single-qubit Hadamard gate can be
simply implemented by a balanced beam splitter transformation
$U_{H}=\left(\frac{1}{\sqrt{2}}\right)^{n_{A}}\exp\left(-\frac{a_{B}^{\dagger}a_{A}}{\sqrt{2}}
\right)\exp\left(\frac{a_{A}^{\dagger}a_{B}}{\sqrt{2}}
\right)\left(\frac{1}{\sqrt{2}}\right)^{-n_{B}}$ working on complete
Hilbert spaces of two modes $A$ and $B$ ($n_{A}, n_{B}$ are photon-number operators).
To test the quality of preparation
of this superposition, we consider another subsequent
Hadamard gate that should reverse the superposition states
$|+\rangle_{AB}$ back to a complementary-basis state
$|1,0\rangle_{AB}=(|+\rangle_{AB}+|-\rangle_{AB})/\sqrt{2}$ .

If the noise photon coming from the environment is
indistinguishable from the signal photon, a coherent
superposition $(\sqrt{2}|0,2\rangle_{AB}+|1,1\rangle_{AB})/\sqrt{3}$ arises
in a larger Hilbert space of the same modes. This superposition can still
perfectly carry any phase information imposed, for example, by
$U_{PS}=\exp(i\phi n_A)$. On the other hand, both the state
$|0,2\rangle_{AB}$ and $|1,1\rangle_{AB}$ generates unavoidable errors
in our implementation of the Hadamard gate
$U_{H}$ due to the presence of another photon. However,
by subtraction of a single
photon from mode B we conditionally reach  state
$(2|0,1\rangle_{AB}+|1,0\rangle_{AB})/\sqrt{5}$ which can be balanced
by simple amplitude damping back to the original superposition
$|+\rangle_{AB}$. It clearly demonstrates that by
combination of classical filtration and quantum noise eater technique,
the original superposition state of a qubit influenced by coherent single-photon noise can
be fully restored.

\section{Theoretical description of experimental test}

\begin{figure} \centerline{\psfig{width=10cm,angle=0,file=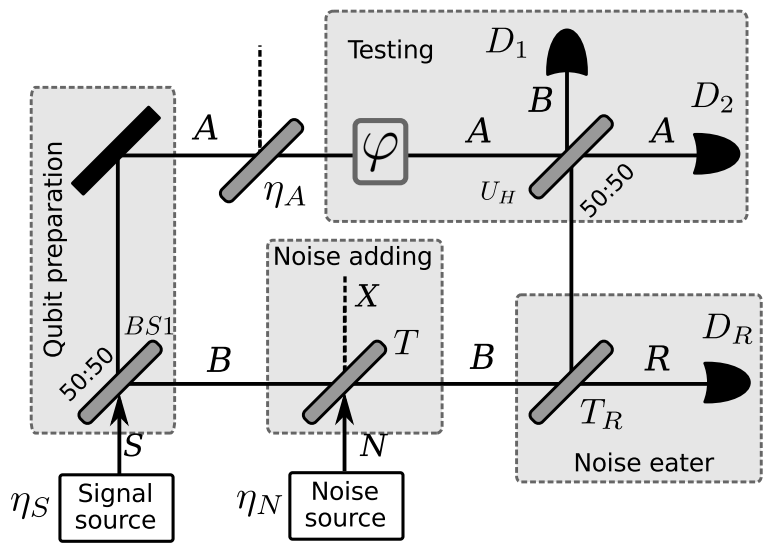}}
\caption{Set-up for proof-of-principle test of coherent noise quantum eater
for dual-rail qubit. $T$ is intensity transmission of the beam splitter
(the $T$ is the transmission ratio of the photons coming from the
left to right), $T_{R}$ is the intensity transmission of the
beam splitter (the $T_{R}$ is the transmission ratio of the photons coming from the
left to right), $D_{i}$ are detectors, $\varphi$ is a phase
shifter, $\eta_{A}$ is the intensity transmission of the beam splitter
(attenuator), signal resp. noise source generate single photon with
probability $\eta_{S}$ resp. $\eta_{N}$.} \label{schema} \end{figure}

A detailed scheme of
the proof-of-principle test of the proposed single-photon noise eater is depicted
in figure \ref{schema}.  The scheme
is divided into stages that help to visualize the whole protocol.
Signal photon enters to the qubit preparation stage from the source of
photons with
probability $\eta_{S}$.
Then a beam splitter BS1 splits the photon equally into both rails, where
the resulting
equatorial dual-rail qubit state is created $\frac{1}{\sqrt{2}}(\ket{0,1}+\ket{1,0})$
  During the noise adding
stage, coherent noise photon is penetrating into the qubit from
single-photon source with probability $\eta_{N}$ by a beam splitter with
intensity transmission $T$ (the $T$ is the transmission ratio of the signal coming from the
left to right in Fig.~1).

To evaluate impact of the single-photon noise on
the qubit without any action of the noise eater, we set $T_{R}=0$.
Then we close the interferometric setup by placing another Hadamard gate
$U_{H}$ that merges both rails.
We can directly quantify the quality of a qubit
after the addition of noise by visibility. To evaluate it we vary phase $\varphi$
in the testing stage and
measure the probability $P(\varphi)$ of a count at the one output port of
the Hadamard gate. Visibility is defined by $V
=\frac{P_{\max}-P_{\min}}{P_{\max}+P_{\min}}$, where
$P_{\max}=\max_{\varphi}P(\varphi)$ ($P_{\min}=\min_{\varphi}P(\varphi)$)
is maximum (minimum) of $P(\varphi)$ over $\varphi$. To maximize
visibility, we use another amplitude damping operations on qubit
represented by an attenuator with intensity transmission $\eta_A$ (transmission ratio of the signal
coming from the left to right in Fig.~1) placed in the rail
$A$.

The probability to detect at least one photon at detector
D$_{1}$ (or D$_2$) with detector efficiency $\eta_D$ will be sum
of three independent terms coming from different states going through
the MZ interferometer. The first state
$|\psi_{1}\rangle=\sqrt{1-\eta_S}\sqrt{\eta_N}|01\rangle_{SN}$ represents the situation of
one photon in noise mode and no photon in signal mode, second state
$|\psi_{2}\rangle=\sqrt{\eta_S}\sqrt{1-\eta_N}|10\rangle_{SN}$ represents the situation of
one photon in signal mode and no photon in noise mode and the third state
$|\psi_{3}\rangle=\sqrt{\eta_S}\sqrt{\eta_N}|11\rangle_{SN}$ represents the situation of
one photon in signal mode and one photon in noise mode.

In the following we show steps of calculation for getting output states
$|\psi_{i}'\rangle$ from the
input states $|\psi_{i}\rangle$. The quantum modes used in the calculations
are shown in the Fig.~\ref{schema}.
The symbols over the right arrows
represent the transformations used to obtain next step,
e.g. $BS(T)$ means beam splitter $T$ with intensity transition $T$,
$PS(\varphi)$ is
a phase shifter with phase shift $\varphi$.

\begin{multline}
  |\psi_{1}\rangle \xrightarrow{BS(T)}
  \sqrt{1-\eta_S}\sqrt{\eta_N}(\sqrt{T}|01\rangle_{BX}+i\sqrt{1-T}|10\rangle_{BX}\xrightarrow{BS(1/2)}
\\
|\psi_{1}'\rangle = \sqrt{1-\eta_S}\sqrt{\eta_N}(\sqrt{T}|001\rangle_{ABX}+i\sqrt{\frac{(1-T)}{2}}(i|100\rangle_{ABX}+|010\rangle_{ABX})
\end{multline}

The probability to detect a photon in mode $B$ with detector efficiency
$\eta_D$ reads $$P_{1}= \Tr(|1\rangle_{B}\langle 1 |\psi_{1}'\rangle\langle \psi_{1}'|)=\frac{\eta_{D}\eta_{N}(1-\eta_S)(1-T)}{2}$$

\begin{multline}
  |\psi_{2}\rangle \xrightarrow{BS(1/2)}
\frac{\sqrt{\eta_S}\sqrt{1-\eta_N}}{\sqrt{2}}(|10\rangle_{AB}+i|01\rangle_{BX})
\xrightarrow{BS(\eta_A),\;  BS(T),\; PS(\varphi)} \\
\frac{\sqrt{\eta_S}\sqrt{1-\eta_N}}{\sqrt{2}}(\sqrt{\eta_A}\exp{(i
\varphi)}|100\rangle_{ABX}+i|0\rangle_{A}(\sqrt{T}|10\rangle_{BX}+\sqrt{1-T}|01\rangle_{BX})\xrightarrow{BS(1/2)}
\\
|\psi_{2}'\rangle = \frac{\sqrt{\eta_S}\sqrt{1-\eta_N}}{\sqrt{2}}(\exp{(i
\varphi)}\frac{\sqrt{\eta_{A}}}{\sqrt{2}}(|10\rangle_{AB}
+i|01\rangle_{AB})|0\rangle_{X} + \\ +
\frac{i\sqrt{T}}{\sqrt{2}}(|01\rangle_{AB} + i|10\rangle_{AB})|0\rangle_{X}
+ i\sqrt{1-T}|001\rangle_{ABX})
\end{multline}

The probability to detect a photon in mode $B$ with detector efficiency
$\eta_D$ reads $$P_{2}= \Tr(|1\rangle_{B}\langle 1
|\psi_{2}'\rangle\langle \psi_{2}'|)=\frac{\eta_{D}\eta_{S}(1-\eta_N)(\eta_A + T
-2\sqrt{\eta_{A}T}\cos\varphi)}{4}$$

\begin{multline}
  |\psi_{3}\rangle \xrightarrow{BS(1/2),\; BS(\eta_A),\; PS(\varphi)}
\frac{\sqrt{\eta_S\eta_N}}{\sqrt{2}}(i|011\rangle_{ABN}+\exp{(i
\varphi)}\sqrt{\eta_{A}}|101\rangle_{ABN})\xrightarrow{BS(T)} \\
\frac{\sqrt{\eta_S\eta_N}}{\sqrt{2}}(i|0\rangle_{A}(i\sqrt{2T(1-T)}|20\rangle_{BX}+
i\sqrt{2T(1-T)}|02\rangle_{BX}+(2T-1)|11\rangle_{BX})+ \\ +
\exp{(i\varphi)}\sqrt{\eta_{A}}|1\rangle_{A}(i\sqrt{1-T}|10\rangle_{BX}+\sqrt{T}|01\rangle_{BX}))\xrightarrow{BS(1/2)}
|\psi_{3}'\rangle = \\
\frac{\sqrt{\eta_S\eta_N}}{\sqrt{2}}(-\sqrt{2T(1-T)}(\frac{1}{2}|02\rangle_{AB}-\frac{1}{2}|20\rangle_{AB}+\frac{i}{\sqrt{2}}
|11\rangle_{AB})+\frac{i(2T-1)}{\sqrt{2}}(|01\rangle_{AB}+|10\rangle_{AB})-
\\ - \exp(i\varphi)\frac{\sqrt{\eta_{A}(1-T)}}
{\sqrt{2}}(|20\rangle_{AB}+|02\rangle_{AB})+\exp(i\varphi)\frac{\sqrt{\eta_{A}T}}{\sqrt{2}}(i|01\rangle_{AB}+|10\rangle_{AB})
\end{multline}

The probability to detect at least one photon (one or two) in mode $B$ with detector efficiency
$\eta_D$ reads
\begin{multline}
  P_{3}= \Tr((|1\rangle_{B}\langle 1|+|2\rangle_{B}\langle 2|)
|\psi_{3}'\rangle\langle \psi_{3}'|)=\\
\frac{1}{4}\left(\eta_{D}\eta_{S}\eta_N(\eta_A(1-\eta_{D}) + 2 +
  T +\eta_{D}T(T-1+\eta_{A}) +
(\eta_{D}(1-T) -1)2\sqrt{\eta_{A}T}\cos(\varphi))\right)
\end{multline}
Adding the three probabilities together gives us
\begin{equation} P = P_{1}+P_{2}+P_{3}
= \frac{\eta_D}{4}\left(W_{1}+W_{2}\cos\varphi\right), \end{equation}
where
$W_{1}=2\,\eta_{{N}}+\eta_{{S}}\eta_{{A}}+\eta_{{S}}\eta_{{N}}
T\eta_D\eta_{{A}}-2\,\eta_{{N}}T-\eta_{{S}}\eta_{{N}}
\eta_DT+\eta_{{S}}T-\eta_{{S}}\eta_{{N}}\eta_D\eta_
{{A}}+\eta_{{S}}\eta_{{N}}{T}^{2}\eta_{D}$ and
$W_{2}=2\,\eta_{{S}}\eta_{{N}}\eta_D\sqrt {T}\sqrt
{\eta_{{A}}}- 2\,\eta_{{S}}\eta_{{N}}{T}^{3/2}\eta_D\sqrt
{\eta_{{A}}}- 2\,\eta_{{S}}\sqrt {T}\sqrt {\eta_{{A}}}$
determine a depth of modulation of the interference fringe. The
corresponding visibility of interference for balanced optical paths
$\eta_{A}=T$ reads
\begin{equation}
  V=\frac{W_2}{W_1}=\frac{\eta_{S}T(\eta_{N}T\eta_{D}-\eta_{N}\eta_{D}+1)}{\eta_{S}\eta_{N}\eta_{D}
  T(T-1)+\eta_{S}T+\eta_{N}(1-T)}.
\end{equation}

It can be further simplified, for equal input signal and noise losses
$\eta_{S}=\eta_{N}$ and for detector efficiency $\eta_D=0.5$ used in the
experiment, to the following form \begin{equation}
V=\frac{2T+\eta_{S}T(T-1)}{2+\eta_{S}T(T-1)}.
\end{equation} For typical $\eta_S=\eta_N\ll 1$, the reduced visibility
simply approaches \begin{equation} V \approx T. \end{equation} Irrespective
  of full coherence between the signal and noise photons,
  visibility is directly proportional to
  the probability that signal photon arrives
  to a detector. The reduction comes simply from the fact that noise photon,
  although fully coherent does not carry information about testing phase
  $\varphi$. Either the signal or noise photon randomly arrives to the
  Hadamard gate and compensation by $\eta_A$ is actually
  redundant. The prepared state remains dominantly in the original
  2D Hilbert space, since contribution of the bunching is
  negligible. For $\eta_S=\eta_N\ll 1$, we get the same visibility $V \approx
  T$ also for noisy photon being fully distinguishable. These two cases are
  therefore problematically distinguishable if only the reduction of visibility
  is analyzed (see figure \ref{graf1} for both plots).

For lower transmission $T$, the reduction of visibility is really
substantial.
 To increase the visibility,
we use an elementary noise eater consisting of beam splitter with intensity
transmission $T_{R}$ (the $T_{R}$ is the transmission ratio of the photons coming from the
left to the right in Fig.~1) and
single photon detector $D_R$ right after the coupling of noise photon.
We optimize $\eta_A$ and $T_R$ and measure the
visibility at the detector $D_{1}$ conditioned now by detection of a photon at
detector $D_{R}$. We exploit the very rare but still present bunching
effect, when the signal and noise photons become indistinguishable. The
conditional probability of photon detection at $D_1$ when one photon has
already been detected
at $D_R$
can be calculated using just the state $|\psi_{3}\rangle$ since the other two
single photon
states $|\psi_{1}\rangle$ and $|\psi_{2}\rangle$ would not contribute to coincidence detection.

We again show the steps of calculations to obtain an output state
$|\psi_{c}\rangle$ from the input state $|\psi_{3}\rangle$.
A $Postselection$ means the we keep just the terms where
mode $R$ contains one photon and mode $A$ or $B$ contains another photon.
\begin{multline}
  |\psi_{3}\rangle \xrightarrow{BS(1/2),\; BS(\eta_A),\; PS(\varphi)}
\frac{\sqrt{\eta_S\eta_N}}{\sqrt{2}}(i|011\rangle_{ABN}+\exp{(i
\varphi)}\sqrt{\eta_{A}}|101\rangle_{ABN})\xrightarrow{BS(T)} \\
\frac{\sqrt{\eta_S\eta_N}}{\sqrt{2}}(i|0\rangle_{A}(i\sqrt{2T(1-T)}|20\rangle_{BX}+
i\sqrt{2T(1-T)}|02\rangle_{BX}+(2T-1)|11\rangle_{BX})+ \\ +
\exp{(i\varphi)}\sqrt{\eta_{A}}|1\rangle_{A}(i\sqrt{1-T}|10\rangle_{BX}+\sqrt{T}|01\rangle_{BX}))\xrightarrow{Postselection}
\\
\frac{\sqrt{\eta_S\eta_N}}{\sqrt{2}}(-2i\sqrt{T(1-T)T_{R}(1-T_{R})}|0110\rangle_{ABRX}+\exp{(i
\varphi)}i\sqrt{\eta_{A}(1-T)T_{R}}|1010\rangle_{ABRX})
= \\
iK_{1}(-K_{2}|0110\rangle_{ABRX}+\exp{(i
\varphi)}K_{3}|1010\rangle_{ABRX})\xrightarrow{BS(1/2)} \\
|\psi_{c}\rangle = i\frac{K_{1}}{\sqrt{2}}(K_{2}(|01\rangle_{AB}+i|10\rangle_{AB})|10\rangle_{RX}+K_{3}\exp(i
\varphi)(|10\rangle_{AB} + i|01\rangle_{AB})|10\rangle_{RX}),
\end{multline}
where we have introduced coefficients
$K_{1}=\sqrt{\frac{\eta_{S}\eta_{N}}{2}}$, $K_{2}=2\sqrt{T(1-T)T_{R}(1-T_{R})}$ and
  $K_{3}=\sqrt{\eta_{A}(1-T)T_{R}}$.
The probability to detect one photon at detector $D_{1}$ (mode $B$) provided
one photon was detected at detector $D_{R}$ (mode $R$)
reads $$ P_{c}=\Tr(|11\rangle_{BR}\langle 11|
\psi_{c}\rangle\langle \psi_{c}|)   = \frac{K_{1}^{2}}{2}\left(K_{2}^2 + K_{3}^2 +
2K_{2}K_{3}\cos\varphi\right)$$
and if we substitute back for $K_{i}$ we get
\begin{multline} P_{c}=
\frac{1}{4}\eta_{S}\eta_{N}T_{R}\eta_D\eta_{R}(1-T)
\left(\eta_{A}+4T(1-T_{R})-4\cos(\varphi)\sqrt{\eta_{A}T(1
-T_{R})}\right).
\end{multline}
For an optimal setting $\eta_{A}=T$ and $T_{R}=3/4$
it gives maximal unit visibility
\begin{equation} V_{\max}=1. \end{equation}
  By the action of the noise eater, the maximal visibility of interference
  is recovered, irrespective of the probability that a noise photon
  appears and irrespective of the values of $\eta_S$, $\eta_N$, $\eta_D$
  and $\eta_R$. We exploit the mutual coherence of signal and noise
  photons and filter out the bunching effect leading to a full recovery of
  the qubit state. It is a role
  which cannot be principally taken by any distinguishable noise photon,
  which gives, after the optimal application of noise eater, threshold
  visibility $V_{\text{th}}=\frac{1}{\sqrt{2}}$ \cite{gav11}. The bunching effect
  affects also selection of the optimal setting, which is established to
  compensate a modulation of amplitude coefficients
  corresponding to quantum operation $aa^{\dagger}$ applied to
  single photon state $|1\rangle$.
  If we are able to observe visibility greater than $V_{\text{th}}$, the noise photon
  had to be at least partially coherent with the signal photon and
  some amount of photon bunching occurred.
  Although this theoretical prediction is very
  promising, the bunching is very subtle and fragile effect, therefore
  a careful experimental test is required to observe it for lower transmission
  $T$ when a substantial reduction of
  visibility appears. The setup itself is an
  interesting combination of single-photon and two-photon interference
  experiments with a direct application to coherent noise reduction.

\section{Experimental realization}

\begin{figure}
  \begin{center}
    \smallskip
     \resizebox{12cm}{!}{\includegraphics*{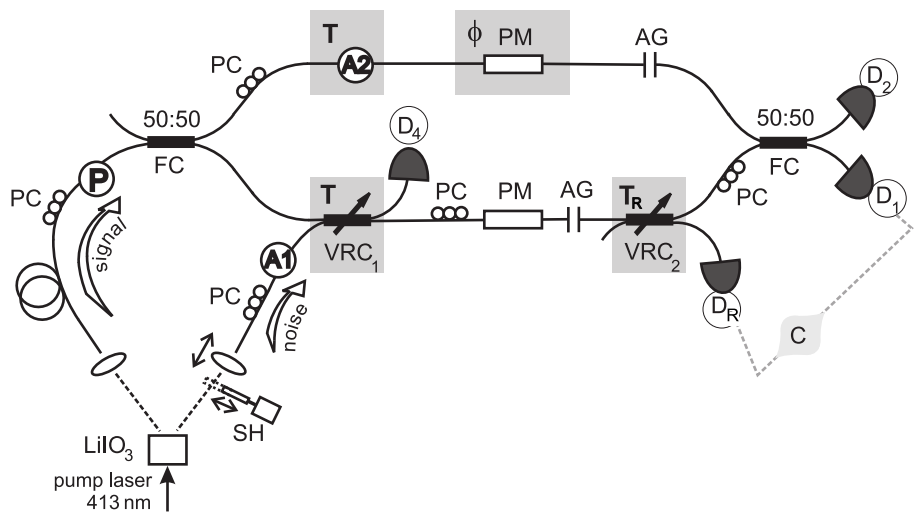}}
    \smallskip
  \end{center}
  \caption{Experimental setup. Shutter (SH), polarization controllers (PC),
  polarizer (P), attenuators (A), phase modulators (PM), adjustable air-gaps (AG),
  fiber couplers (FC), variable-ratio couplers (VRC), detectors (D).}
  \label{setup}
\end{figure}

Our experimental setup (see figure \ref{setup}) was built using
fiber optics that allow us to simply control transmissivity $T$
and $T_{R}$ of the beam splitters via variable-ratio couplers with the
range of transmissivity $0-1$.
The main part of the setup is a balanced Mach-Zehnder (MZ) interferometer.

Signal and noise photons were created by type-I degenerate
spontaneous parametric downconversion in a nonlinear crystal
pumped by a continuous laser (413~nm). Photons from each pair are
tightly correlated in time. They have the same polarization state
and spectrum whose bandwidth is determined mainly by the coupling of
photons from the nonlinear crystal into single-mode fibers.

Before the measurement the source of photon pairs was adjusted by
optimizing a visibility of two-photon interference at the
variable-ratio coupler VRC$_1$ with splitting ratio set to 50:50. The
visibility of Hong-Ou-Mandel (HOM) dip \cite{HOM} typically
reached values about $0.98$. Then the intensities of signal and
noise were balanced. It was
realized by a measurement of count rates at detectors D$_R$ and
D$_4$ while the transmissivity of coupler VRC$_2$ was set to $1$. We
tuned the count rates of noise at these detectors to be twice of
the count rates of the signal.

Single photon visibility behind  the MZ interferometer was above $0.93$.

Measurements were realized with maximally
indistinguishable signal and noise photons, when HOM dip was set to its
minimum. The signal to noise ratio was determined by intensity transmissivity $T$ of
VRC$_1$. Transmissivity of the other arm of MZ interferometer was
also set to the value of $T$. Coupler VRC$_2$ together with detector
D$_{3}$, implemented the noise eater. Its transmissivity $T_{R}$ was not compensated in the
other arm of interferometer.

All used detectors were Perkin-Elmer single-photon counting
modules. To implement
postselection measurements the signals from detectors were processed by
coincidence electronics based on time-to-amplitude converters and
single-channel analyzers. The coincidence window was set to 2~ns
when accidental coincidence rates were negligible.

The phase of light in optical fibers is influenced by temperature
changes and gradients. This undesirable phase drift was reduced by
a thermal isolation of MZ interferometer and by an active
stabilization of phase that was applied before each few-second
measurement step.

First we measured how the visibility of interference at the
outputs of MZ interferometer is damaged by the presence of noise.
During this measurement no postselection was applied and so the
transmissivity $T_{R}$ of VRC$_2$ was set to zero. It means that if
the noise input is shut then the visibility at interferometer
outputs maximal.

For large losses the visibility for indistinguishable and distinguishable
cases degrades to the
same value $V\approx T$.

The aim of this work was to increase the visibility at outputs of
MZ interferometer using a coherent noise eater based on single photon subtraction.
We measured coincidence rate $C_1$ between detectors
D$_1$ and D$_R$. Intensity transmissivity of VRC$_2$ was adjusted
to the value which maximizes the visibility of coincidence rate $C_1$
 $T_{R}=3/4$. The
visibility of $C_1$ was measured as a function of transmissivity $T$.

\section{Experimental results}

\begin{figure}
  \begin{center}
    \smallskip
     \resizebox{15cm}{!}{\includegraphics*{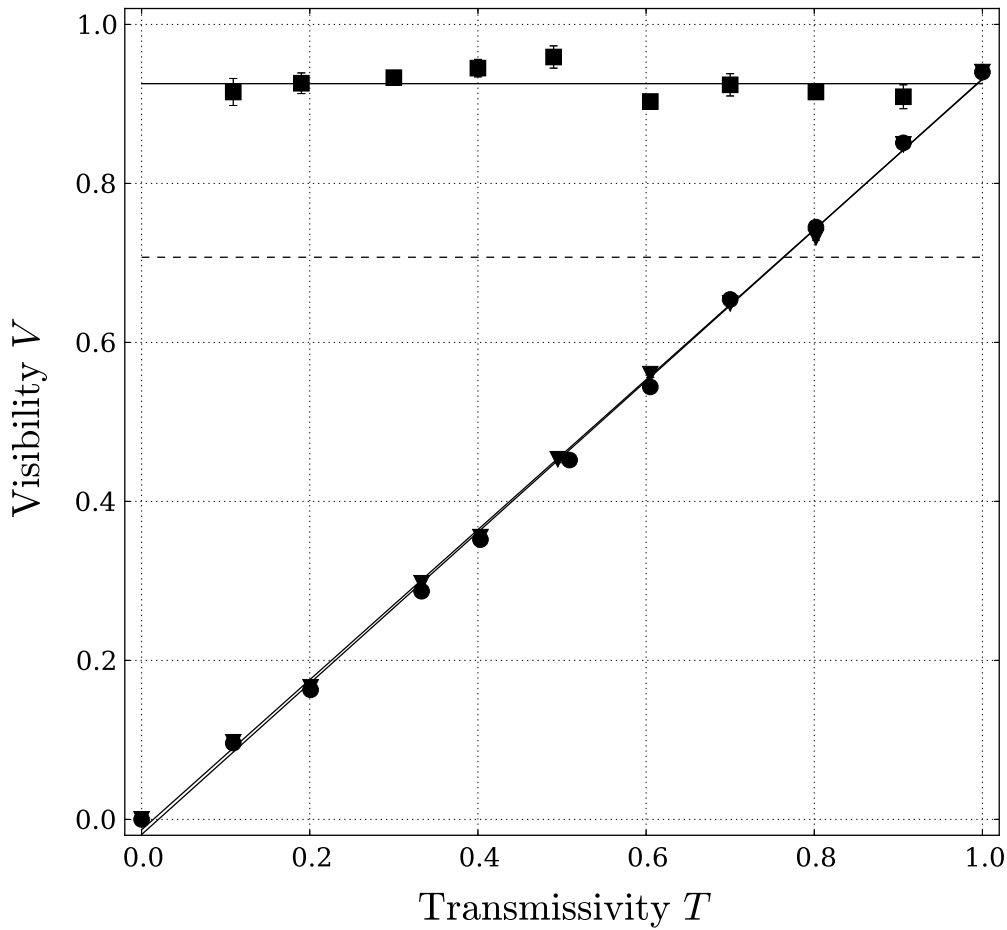}}
    \smallskip
  \end{center}
  \caption{Dependence of visibility on transmissivity $T$.
  Symbols denote experimental results; circles correspond to the visibility at
  MZ interferometer outputs for distinguishable case and triangles for
  indistinguishable case, squares correspond to the visibility after recovery
  for indistinguishable photons. Solid lines are fits to measured data. The
dashed line shows our benchmark value of visibility equal to $1/\sqrt{2}$.}
  \label{graf1}
\end{figure}

\begin{figure}
  \begin{center}
    \smallskip
     \resizebox{15cm}{!}{\includegraphics*{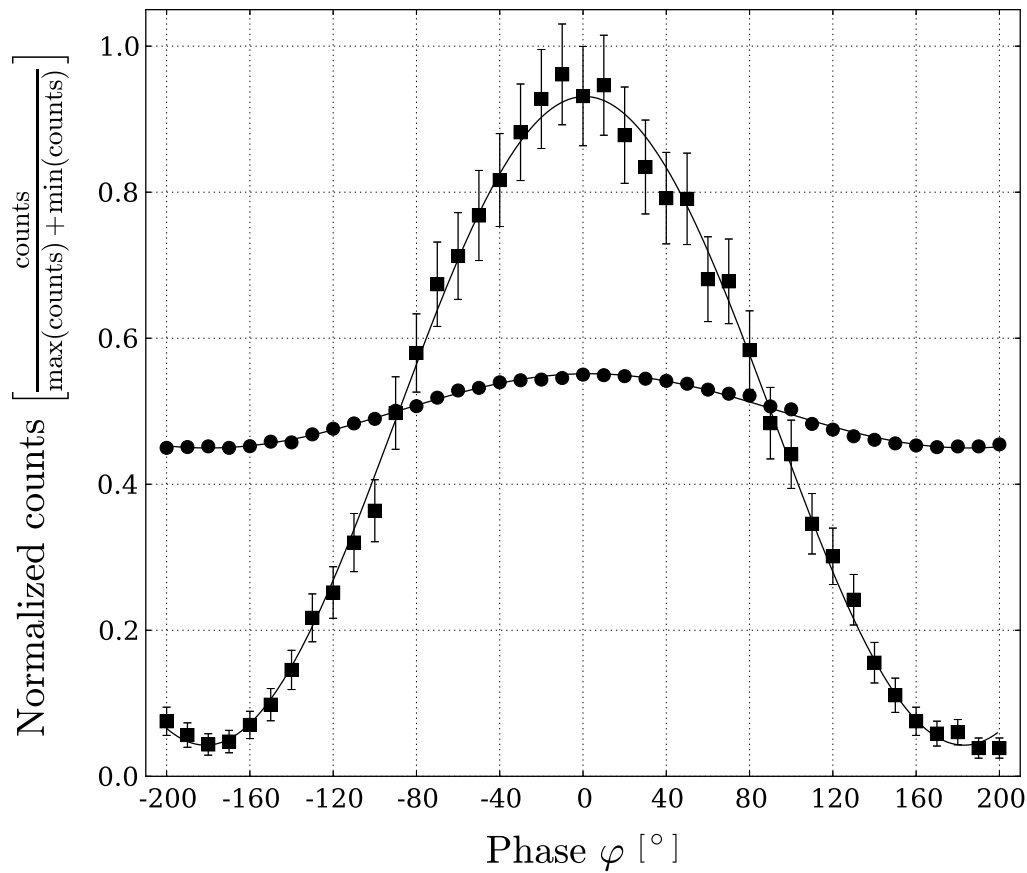}}
    \smallskip
  \end{center}
  \caption{The interference fringes for indistinguishable photons and
    coupling ratio $T=0.109$. Symbols denote experimental results:
  circles correspond to interference fringe without correction and
squares correspond to interference fringe after the recovery. Solid
lines represent fits to measured data.}
  \label{prouzek_ner}
\end{figure}

\begin{figure}
  \begin{center}
    \smallskip
     \resizebox{15cm}{!}{\includegraphics*{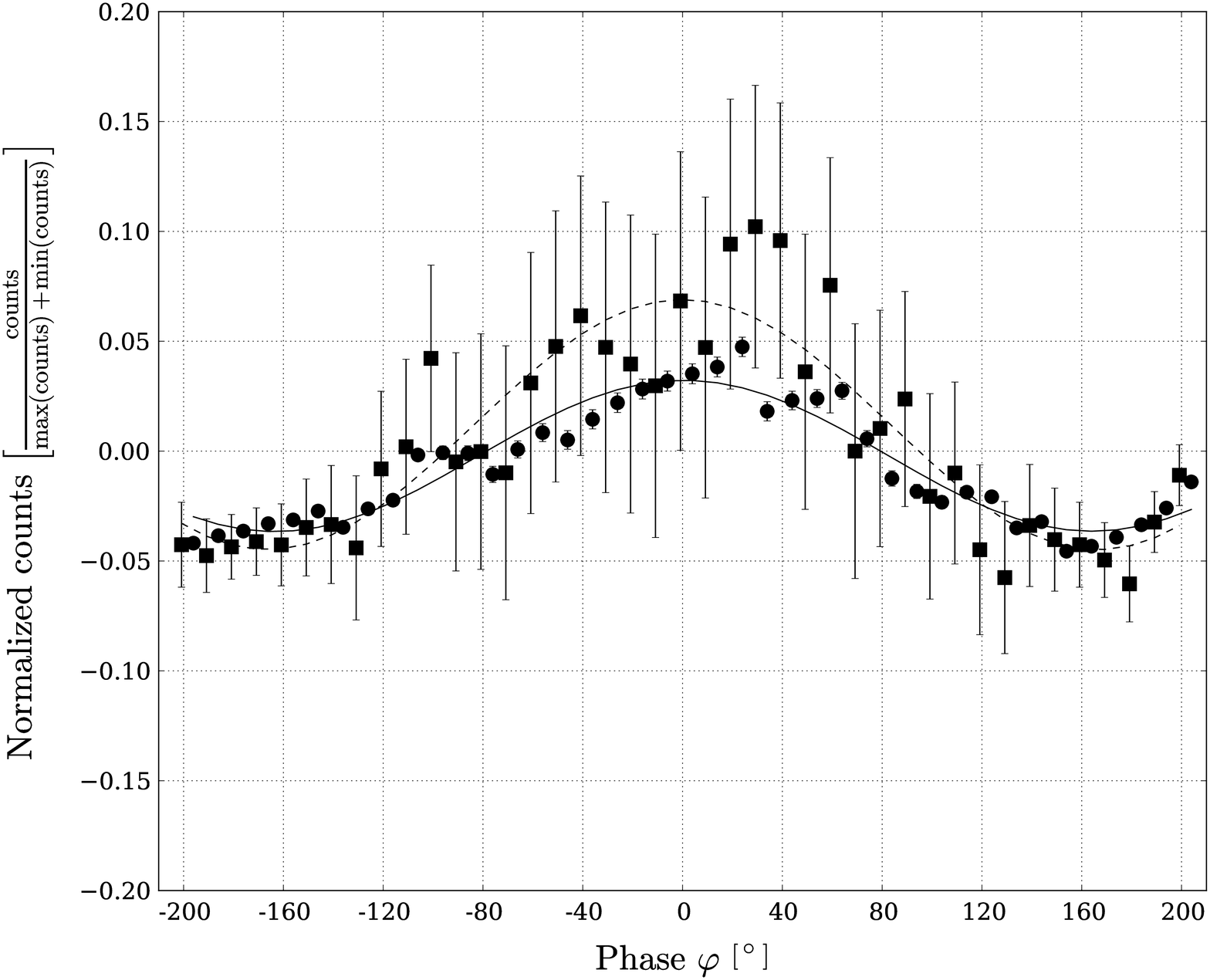}}
    \smallskip
  \end{center}
  \caption{Circles correspond to the difference between the ideal sine interference fringe
    ($V=1$) and the best achieved interference
    fringe for $T=1$ ($V=0.936$); squares correspond to the difference between the
    recovered fringe for $T=0.109$ ($V=0.915$) and the ideal sine fringe ($V=1$). Solid
curve corresponds to the difference between the ideal sine fringe and the fit of the
best achieved interference fringe for $T=1$. Dashed curve corresponds
to the difference
between the ideal sine fringe and the fit of the recovered fringe for
$T=0.109$.}
  \label{prouzky_rozdil}
\end{figure}

Measured visibilities are displayed in figure~\ref{graf1}.
Interference fringes were investigated in the range
of phases $[-120^{\circ},120^{\circ}]$ with a step $10^{\circ}$.
Each point of interference fringe was measured 3-5~seconds,
depending on a quantity of signal. Before each measurement the degree
of phase drift in MZ interferometer was checked and in case of
need it was minimized by a stabilization procedure.  An
interference fringe was measured several times. We added all these
results together and then fitted data. Shown error bars are given
by the Poisson distribution of photo-count statistics.

Visibility of signals at the outputs of MZ interferometer are
influenced by dark counts of detectors. Hence we subtracted the
minimum of corresponding dark count rates from measured count
rates. Final visibilities as a functions of $T$ were fitted by the
curve $a\cdot T+b$ with two parameters $a$ and $b$. Of course,
parameter $a=0.936$ corresponds to the value of visibility at
point $T=1$.

The other visibilities were measured in coincidence measurements
and therefore no correction was needed. The visibility of
coincidence rate $C_1$ does not depend on $T$. Obtained mean value
of visibility is
$0.924\pm0.018$
 (the theoretical value is 1).
One can see that the fit is sometimes out of range of the error bars. It is caused by the
fact that the precision of the measurement results
depends on the accuracy of setting of minimum of HOM dip
and on the fluctuation of this position during the measurement.
The threshold value of visibility, $1/\sqrt{2}$, is plotted by a
dashed line.

We have plotted an interference fringe (number of counts as
a function of phase $\varphi$) for $T=0.109$ without and with
recovery in the Fig.~\ref{prouzek_ner}. To compare
single photon counts and coincidence counts in one figure
we have normalized them by sum of maximum counts and minimum counts
measured in the range of the interference fringe.
The less modulated fringe, plotted by circles, corresponds to the
interference fringe measured without recovery. Visibility obtained from the fit is
$0.097\pm 0.005$. If the  noise eater is switched on the interference fringe becomes
more modulated as is demonstrated by a curve with square symbols.
The visibility increased to the value of $0.915\pm 0.017$.
Shown error bars are given
by Poisson distribution of photo-count statistics.
The error bars significantly increase in case of a coincidence measurement.
It is a consequence of the fact
that the number of total counts in the coincidence detection is less then
in single photon detection by two orders of magnitude in average.
on average two orders less number of total counts in
the coincidence detection.

In Fig.~\ref{prouzky_rozdil} we compare difference between measured and
ideal interference fringes.
First curve denoted by circles is a difference between an ideal sine fringe
and a best achieved interference fringe for $T=1$ without the recovery
($V=0.936$).
The fit to the data points is plotted by a solid line.
The other curve denoted by squares corresponds to a
difference between an ideal sine fringe and a best achieved fringe for
$T=0.109$ with the recovery ($V=0.915$). The fit to the data points is
plotted by a dashed
line.
The plots show that a difference between the actual fringe and the ideal
sine fringe is slightly modulated.
The largest differences are in maxima and minima of interference fringes.
There is a greater difference between the ideal sine
fringe and the fringe with the recovery than the fringe without the recovery.
Also the error bars are much greater in the case of the fringe with
the recovery, being again the consequence of a coincidence measurement.

\section{Conclusion}

We have proposed and experimentally demonstrated an elementary quantum
noise eater for dual-rail qubit influenced by a randomly arriving coherent
photon. The superposition of basis states carried by qubit is changed by
that coherent photon and subsequently, visibility of interference of a
dual-rail
qubit behind the Hadamard gate is decreased. Theoretically, a perfect
recovery of the superposition carried by the qubit has been predicted for
the case
when noise eater is applied after a coherent noise photon was randomly added.
Experimentally, after the recovery we observe visibility
 $V=0.915$ for $T=0.109$, comparing to $V=0.097$ before the recovery,
 that is eight-fold increase of the visibility.
 The value $V=0.915$ is 12 standard deviations
 above the threshold, $V_{\text{th}}=1/\sqrt{2}$, for coherent noise
impacts \cite{gav11}.
 It is the first proof-of-principle
test of a general quantum method of coherent noise eater for qubits which
can protect qubits against coherent noise by
a partial nondestructive and coherent selection of given number of photons.
 A weak temporal and spatial
coherence of signal and noise can be enhanced by fully classical spectral
filtering, as have been many times demonstrated \cite{soub03,belli03,abol09}.
It would be an
interesting extension of this experiment to combine it with an induction of
coherence between the qubit and noise photons.
The scheme of the noise eater can
also be extended to coherent noisy particle disturbing both paths of the
interferometer. Required joint particle subtraction not distinguishing the
paths can be implemented.

\section*{Corollary}

\begin{figure}
  \begin{center}
    \smallskip
     \resizebox{15cm}{!}{\includegraphics*{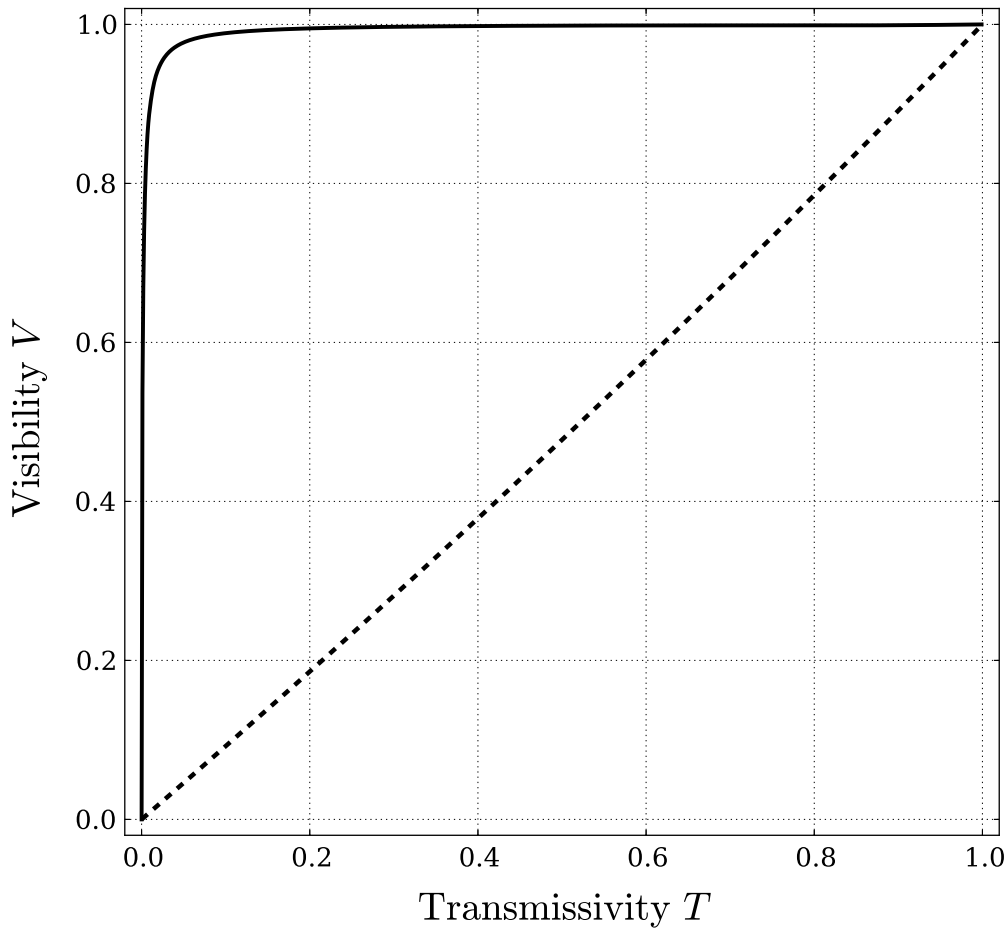}}
    \smallskip
  \end{center}
  \caption{Numerical simulation of the noise eater when also two-photon
    noise contributions are present. The parameters are
    $\eta_{S}=\eta_{N}=0.001$, $\epsilon=0.05$ and $T_{R}$ has been
    optimized for every value of $T$.
    Dashed line represents the situation without the action of the
  noise eater. Solid line represents the situation with the action of the
noise eater.}
  \label{pokorekci}
\end{figure}

We have also studied the action of the noise eater in a more realistic
situation when noise contained more than one photon in a Fock state.
We have numerically simulated behaviour of the noise eater when the signal
was represented by single photon state $|1\rangle$ and
to the noise port of the Mach-Zehnder interferometer was injected
a state $p_0 |0\rangle + p_1 |1\rangle + p_2 |2\rangle$.
We have chosen the coefficients $p_i$ to represent the Poisson
statistics in order to mimic a weak coherent state. The average number of
photons in the noise mode reads $\lambda^2 e^{-2\lambda}+(1/2)\lambda^4
e^{-2\lambda}$, where we have used $p_{k}=\lambda^{k} e^{-\lambda}/k!$
Such a statistical model also very well corresponds to our experimental
situation where we exploit the SPDC process in nonlinear crystal to create both the signal and
noise states of photons.  If we take into account also four-photon pairs created
during the process our output state from SPDC for $\epsilon \ll 1$  reads $(1-\epsilon^{2}/2)|00\rangle + \epsilon|11\rangle +
\epsilon^{2}|22\rangle$. The mean photon number generated locally in the
noise mode is equal to $\epsilon^2 + 2\epsilon^4$. Given $\epsilon$ we
can calculate the corresponding $\lambda$. We have numerically tested action
of the noise eater on both states for different values of related parameters $\epsilon$ and
$\lambda$. The simulation give the same results for both corresponding
states.
In the Fig.~\ref{pokorekci} we have plotted
a result of numerical simulation for $\eta_S=\eta_N=0.001$ and $\epsilon=0.05$, which corresponds to
$0.003$ average number of photons.
The $T_{R}$ had to be optimized for every chosen $T$.
The dashed line
represents the visibility without action of the noise eater
 and the solid line with the action of the noise eater. Due to
the presence of the two-photon events in the noise mode both curves are moved to
lower values compare to single photon noise scenario studied in previous
sections. But there is still a substantial  improvement of the visibility after the action
of the noise eater.

\section*{Acknowledgements}

L.\v{C}. thanks to Jan Soubusta for his help with the experiment. R.F. and M.G.
acknowledge the project GA 205/12/0577 of GA\v{C}R. M.D. acknowledges the
projects PrF-2012-019 and PrF-2013-008 of the Palacky University.



\end{document}